# Fault-Tolerant Routing in Hypercube Networks by Avoiding Faulty Nodes

Shadrokh Samavi, Pejman Khadivi
Department of Electrical and Computer Engineering
Isfahan University of Technology,
Isfahan, Iran 84156-83111

*Abstract*: Next to the high performance, the essential feature of the multiprocessor systems is their fault-tolerant capability. In this regard, fault-tolerant interconnection networks and especially fault-tolerant routing methods are crucial parts of these systems. Hypercube is a popular interconnection network that is used in many multiprocessors. There are several suggested practices for fault tolerant routing in these systems. In this paper, a neural routing method is introduced which is named as Fault Avoidance Routing (FAR). This method keeps the message as far from the faulty nodes as possible. The proposed method employs the Hopfield neural network. In comparison with other neural routing methods, FAR requires a small number of neurons. The simulation results show that FAR has excellent performance in larger interconnection networks and networks with a high density of faulty nodes.

**KEYWORDS**: Hypercube, interconnection networks; fault tolerance; routing; artificial neural networks.

## 1. INTRODUCTION

Networking has numerous application in different fields, such as mobile communications and visual sensor structures [1-2]. In multiprocessor systems networks play essential role in the message passing activities and affect the performance of these systems. Hypercube and Mesh are popular interconnection networks used in some of the message-passing systems such as nCube/10, iPSC/2 and the MIT J-Machine. In the literature, there are some fault-tolerant routing schemes for the Mesh interconnection networks. In this paper, we introduce a new fault-tolerant routing method for a hypercube structure using artificial neural networks.

Figure 1 shows a four-dimensional hypercube. An $n-dimension$ hypercube is an interconnection network with $2^n$ nodes, where each node is addressed with an n-bit string. Two nodes, such as X and Y, are neighbors in the $i$-th dimension, if and only if their addresses differ only the i-th bit. Therefore, if X and Y are defined in the following manner:

$$X = (x_{n-1},...,x_i,...,x_1,x_0) \quad ; \quad x_j \in \{0,1\} \quad (1)$$

$$Y = (y_{n-1},...,y_i,...,y_1,y_0) \quad ; \quad y_j \in \{0,1\}$$

Then, X and Y are neighbors in the i-th dimension if and only if the following conditions hold:

$$y_j = x_j \quad for \quad i \ne j$$
$$y_j = 1 - x_j \quad for \quad i = j \quad (2)$$

By increasing the number of nodes and the complexity of the multiprocessor systems, the need for a fault tolerant design of these systems is more pronounced. Designing fault tolerant interconnection networks have been done in various methods. Inserting spare nodes and channels and reconfiguration under fault conditions is one of these techniques [5-6]. Network embedding is another technique that is used for this purpose [7-9]. A third method is the employment of a fault-tolerant routing algorithm.

For this reason, there are different methods for different kinds of interconnection networks. Most of the algorithms, which are mentioned in the literature, use local information. The knowledge of the state of neighboring nodes is referred to as the *local information*. A fault-tolerant routing algorithm is one that could find a fault-free path between a fault-free source and destination, at the presence of faulty nodes or channels.

There are several classical fault-tolerant routing algorithms designed for the hypercube which most of them use the local information. The routing method in [4] requires each node to know every faulty node within its k-hop distance, and it tries to find a path with no faulty nodes in the next k-hops. However, the search for a path is a difficult task since a complete search requires analysis of k! paths. Also, in [4], another method is introduced which applies the concept of an unsafe node by using the local information. There is yet another method, introduced in [5], which tries to route a message using fault free channels that get the message closer to the destination. In [6], similar to

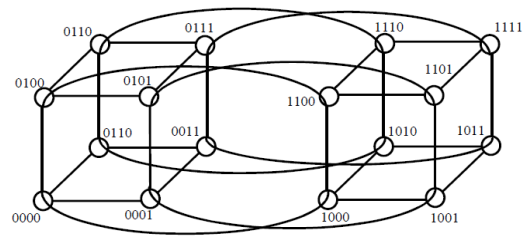

Figure 1. A 4-dimension hypercube

[4], the concept of an unsafe node is used, but the decision-making is more intelligent. In [7] the idea of *routing capability* is introduced. It is defined that a node has routing capability $k$ if it can communicate with all nodes within its k-hop distance. Therefore, it is better to choose nodes with higher routing capability.

The idea of routing using neural networks was first introduced by Winarsk and Raush in 1988 [8]. The purpose of most of the neural routing methods, which are presented so far, is to balance the traffic in the networks with irregular topology. These methods, usually, use a large number of neurons. This number is in the order of $n^2$, where $n$ is the number of nodes in the interconnection network [9-12]. With the particular features of the multiprocessor systems, such as the regular topology of the interconnection networks, which is used in these systems, it is expected that the neural routing methods require a small number of neurons [2].

In the following, a classical routing method is discussed. After that, the Fault Avoidance Routing (FAR) method is introduced. Finally, the simulation results are conferred.

## 2. A CLASSICAL ROUTING METHOD

The routing methods discussed by Lee [4] and Chiu [6], are based on the concept of unsafe nodes. An unsafe node is an entrance to a faulty region of the interconnection network. Not sending the messages to these nodes could result in relatively smaller average path length.

Based on what is discussed by Chiu, the unsafe node is one that has 2 or more faulty or 3 or more unsafe neighbors [6]. According to Lee, an unsafe node has 2 or more faulty or unsafe neighbors [4]. Since using the Chiu's definition of an unsafe node results in lower number of unsafe nodes in a Hypercube, we pick the Chiu's definition in our proposed routing method. The unsafe nodes are categorized into two groups of *strongly unsafe* and *ordinary unsafe* nodes. A *strongly unsafe node* is a node that all of its neighbors are faulty or unsafe. If an unsafe node has one or more safe nodes among its neighbors, then that node is an *ordinary unsafe node*. The set of faulty, safe, unsafe, ordinary unsafe and strongly unsafe nodes are represented by $F$, $S$, $U$, $\overline{U}$, and $\widetilde{U}$ respectively. If $n$, is any node of the interconnection network, then, $N(n)$ is the set of the neighbors' of $n$ and $D(n,t)$ is the set of the neighbors that make the message closer to the destination. Chiu's algorithm prefers the ordinary unsafe node to a strongly unsafe one when it comes to deciding between one of the two. Chiu's algorithm is shown in Figure 2. There, $c$ is the current node and $t$ is the destination node. Also, $d(c,t)$ is a function which returns the distance between nodes $c$ and $t$. Based on this algorithm, safe neighbors that make the message closer to the destination are preferred over any other ones. Next best option, are ordinary unsafe nodes that make the message closer to the target.

```
Algorithm Chiu(c,t)
{ c is the current node address , t is the target node
address}
BEGIN
h=d(c,t);
IF h=0 THEN deliver the message to node c and exit
ELSE IF ∃n ∈ D(c, t) ∩ S
        THEN next_node=n
ELSE IF ∃n ∈ D(c, t) ∩ U
        THEN next_node=n
ELSE IF ∃n ∈ D(c, t) ∩ Ũ  AND ( c ∈ Ũ  OR h ≤ 2 )
        THEN next_node=n
ELSE IF ∃n ∈ ( N(c) − D(c, t)) ∩ S
        THEN next_node=n
ELSE IF ∃n ∈ ( N(c) − D(c, t)) ∩ U
        THEN next_node=n
ELSE    "Error in delivery"
END;
```

Figure 2. Chiu's algorithm

## 3. FAR: FAULT AVOIDANCE ROUTING

In this section, a new routing method using neural networks is introduced. FAR aims to send a message to a neighbor that, first of all, is closer to the destination and secondly, keeps it as far from the faulty nodes as possible. Each node in an n-dimension hypercube has exactly n neighbors. For any neighbor, $i$, in the jth dimension, a cost function, $G_i(j)$, is considered. First, let's define $d(x, y)$ as the Hamming distance between two nodes such as x and y in a hypercube:

$$d(x, y) = \sum_{i=1}^{n} x_i \oplus y_i \qquad (3)$$

where $x_i$ and $y_i$ are the i-th bit of the addresses of the two nodes. Also, suppose that the fault information is in a bit-vector, $F$, with $2^n$ one bit elements, where each element is such that:

$$f_i = \begin{cases} 1 & , if\ node\ i\ is\ faulty \\ 0 & , Otherwise \end{cases} \qquad (4)$$

Hence, if i, j N is the neighbor of a node $i$ in dimension $j$, and $i$ is the current node and $D$ is the destination node, then we define:

$$G_i(j) = d(N_{i,j}, D) + \sum_{k=1}^{2^n} \frac{f_k}{d(N_{i,j}, k) + \varepsilon} \qquad (5)$$

where $\varepsilon$ has a small value. For the implementation purposes, $\varepsilon \neq 0$ prevents the cost function from diverging, but in theory, $\varepsilon$ could be zero. In equation (5), the first term is the distance between j-dimension neighbor of node $i$, and the destination. The second term gives the inverse of the sum of the distances between $N_{i,j}$ and faulty nodes in the interconnection network. Any $N_{i,j}$ that minimizes $G_i(j)$ is a better preference for routing. Therefore, after the cost function is computed for every neighbor, the one with the smallest cost is selected.

To implement FAR using the neural networks, one neuron is assigned for each neighbor. When neural network converges, neuron relative to the neighbor with less cost must be *on*. Therefore, the energy function of the neural network has two parts. The first part implies that the neuron corresponding to the neighbor with the smallest cost must be on. Hence:

$$E1 = (\sum_{j=1}^{n} G_i(j)V_j)^2 \quad (6)$$

where $V_j$ is the output of the neuron assigned to the neighbor on the *j-th*-dimension. The second part of the energy function implies that only one neuron must be on. Hence:

$$E2 = (\sum_{j=1}^{n} V_j - 1)^2 \quad (7)$$

Combining (6) and (7) we have:

$$E = K_1 E1 + K_2 E2 = K_1(\sum_{j=1}^{n} G_i(j)V_j)^2 + K_2(\sum_{j=1}^{n} V_j - 1)^2 \quad (8)$$

where $K_1$ and $K_2$ are constant coefficients. We decided to use the continuous-time Hopfield network. The motion equation for this neural network is [13]:

$$\frac{dU_i}{dt} = \sum_{j=1}^{n} W_{ij} V_j + T_i \quad (9)$$

and :

$$\frac{\partial E}{\partial V_i} = -\frac{dU_i}{dt} \quad (10)$$

where $U_i$ is the input of neuron $i$, $W_{ij}$ is the weight between neurons $i$ and $j$, and $T_i$ is the threshold of neuron $i$. Taking the partial derivative of (8) concerning $V_i$ would result:

$$\frac{\partial E}{\partial V_j} = 2K_1 G_i(j)(\sum_k G_i(k)V_k) + 2K_2(\sum_k V_k - 1) \quad (11)$$
$$= -2K_2 + \sum_k (2K_1 G_i(j)G_i(k) + 2K_2)V_k$$

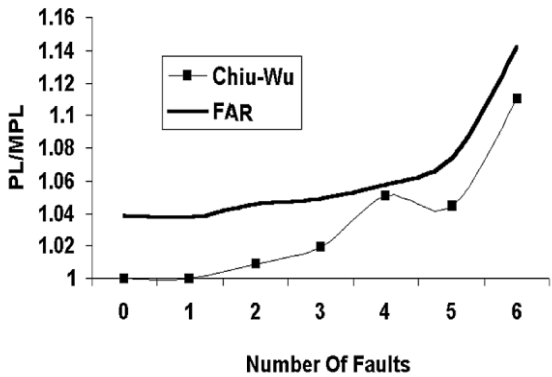

Figure 3. Comparison of FAR and the Chiu's method for a 4-dimensional network

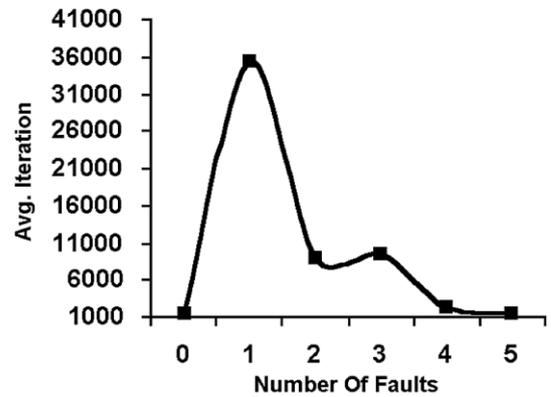

Figure 5. FAR's average number of iterations in a four dimensional network

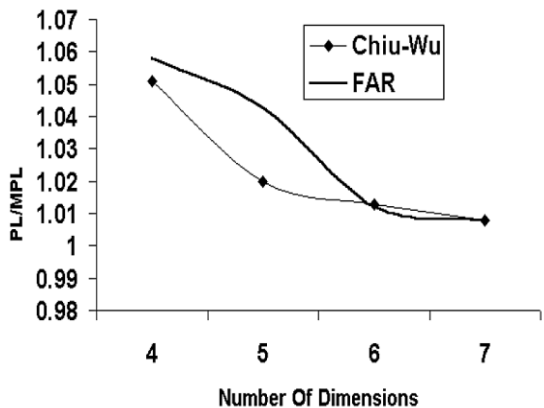

Figure 4. Comparison of FAR and the Chiu's algorithm in different networks with 4 faults

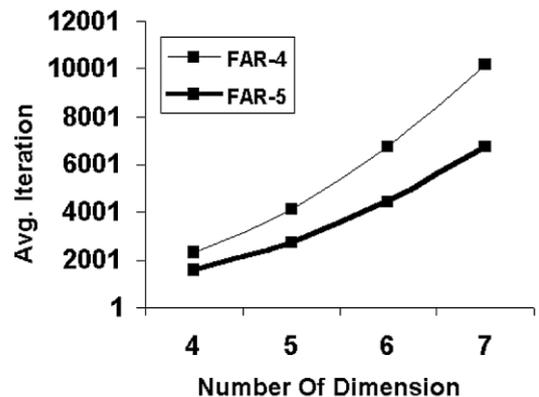

Figure 6. FAR's average number of iterations for networks with 4 and 5 faulty nodes

Using (9), (10) and (11), one can find the weights and the threshold values of a continuous-time Hopfield neural network for the FAR method:

$$\begin{cases} T_j = 2K_2 \\ W_{jk} = -(2K_1 G_I(j)G_I(k) + 2K_2) \end{cases} \quad (12)$$

Therefore, the FAR method for an n-dimensional hypercube interconnection network, can be implemented using only n neurons.
To achieve a higher degree of freedom, equation (5) could be rewritten as below:

$$G_i(j) = K_3 d(N_{i,j}, D) + K_4 \sum_{k=1}^{2^n} \frac{f_k}{d(N_{i,j}, k) + \epsilon} \quad (13)$$

where K3 and K4 are constant coefficients.

## 4. SIMULATION RESULTS

In this section, the results obtained from the simulation of the Chiu's method and that of the FAR are discussed. For the simulation of the neural FAR method to compute $G_i(j)$ the equation (13) is employed. The simulator used the following coefficients:
$K_4 = 0.42$, $K_3 = 1$, $K_2 = 15$, $K_1 = 0.01$
It should be noticed that the relatively small value of $K_1$ compared to $K_2$ is because that in the energy function of the FAR method, the term with $K_1$ is larger than that of the $K_2$. All of the mentioned coefficients are found by trial and error.
In regards to the simulation process, we refer to a case as a fixed size network with a fixed number of faulty nodes. Then, for every situation, there are some runs. For every run, the location of each faulty node is randomly selected. Also, the addresses of the fault-free source and destination are randomly chosen. After a large enough number of runs, the mean path length, MPL, for every case is then computed. Comparing MPL of one case with that of another case would be misleading. This because the average Hamming distance for each case is different. Therefore, MPL for each case is normalized by the fault-free mean path length, producing the PL/MPL parameter.
Figures (3) and (4) compare the FAR method with Chiu's algorithm. It can be seen that, for small networks, Chiu's method performs better. But with the increase in the size of the networks and number of the faulty nodes, the performance of FAR approach to that of Chiu's.
It is expected that with further increase in the dimensions of the network and density of the faults, FAR turns out to be superior. In these situations, Chiu's performance approach to that of an algorithm that uses only local information, while FAR, even though indirectly, enjoys global information. When the network is small, the indirect use of global fault information, similar to what FAR does, could mislead the system. But with an increase in the number of nodes, the global information that comes from all the faulty nodes can help the routing procedure.
Figures (5) and (6) show variations in the number of iterations for the FAR method.

## 5. CONCLUSIONS:

In this paper, a new fault-tolerant routing algorithm is introduced that uses neural techniques. The main mechanism of the suggested method is to keep the message as far from the faulty nodes as possible. Our approach has two advantages over other neural methods: one is employing a small number of neurons and the second one is the considering of the faults in a hypercube network. For large networks, our suggested method shows a shorter average path length compared to the classical routing methods. This is due to applying global information in the computations. The proposed method is also under consideration for routing of remotely guided robots in a warehouse environment with a mesh network of sensors.

## REFERENCES:

[1] Khadivi, P., Samavi, S., and Todd, T.D., "Multi-constraint QoS routing using a new single mixed metrics." Journal of Network and Computer Applications 31, no. 4, 2008, pp. 656-676.
[2] Hooshmand, M., S.M.R. Soroushmehr, P. Khadivi, Samavi, S., and Shirani, S., "Visual sensor network lifetime maximization by prioritized scheduling of nodes." Journal of Network and Computer Applications 36, no. 1, 2013, pp. 409-419.
[3] Rowley, R.A. and Bose, B., "Distributed Ring Embedding in Faulty De Bruijn Networks," *IEEE Transactions on Computers*, Volume 46, Number 2, 1997, pp. 187-190.
[4] Lee, T.C., and Hayes, J.P., "A Fault-Tolerant Communication Scheme for Hypercube Computers,"
*IEEE Trans. On Computers*, Vol. 41, No. 10, 1992, pp. 1242-1256.
[5] Loan, Y., "A Fault-Tolerant Routing Algorithm In Hypercubes," *Proc. Of International Conference On Parallel Processing*, 1994, pp. 163-166.
[6] Chiu, G.M., and Wu, S.P., "A Fault-Tolerant Routing Strategy in Hypercube Multicomputers," *IEEE Trans. On Computers*, Vol.45, No.2, 1996, pp. 143-155.
[7] Chiu, G.M., and Vhen, K.S., "Use Of Routing Capability For Fault-Tolerant Routing In Hypercube Muticomputers," *IEEE Trans. On Computers*, Vol.46, No. 8, 1997, pp 953-958.
[8] Rauch, H.E., and Winarske, T., "Neural Networks For Routing Communications Traffic," *IEEE Cont. Syst. Mag.*, 1988, pp 26-31.
[9] Kamoun, F., and Mehmet Ali, M.K., "A Neural Network Shortest Path Algorithm For Optimum Routing In Packet-Switched Communication Networks," *Global Telecommunication Conference*, Vol.1, 1991, pp. 120-124.
[10] Pierre, S., and Said, H., and Probst, W.G., "A Neural Network Approach For Routing In Computer Networks," *IEEE Canadian Conference On Electrical and Computer Engineering*, Vol. 2, 1998, pp. 826-829.
[11] Wang, J., "A Dual Neural Network For Shortest Path Routing", *International Conference On Neural Network*, Vol. 2, 1997, pp. 1295-1298.
[12] Baba, T., and Funabili, N., "A Routing Algorithm Based On Greedy Neural Network For Non-Sequential Activation Scheduling Problems", *IEEE International Conference on Systems, Man, and Cybernetics*, Vol.3, 1998, pp. 2285-2290.
[13] Takefuji, Y., *Neural Network Parallel Computing*, Kluwer Academic Publishers, 1992.